\documentclass[reprint,secnumarabic,amssymb, nobibnotes, nofootinbib, aps, prd, onecolumn]{revtex4-2}
\usepackage[utf8]{inputenc}
\usepackage{xcolor}
\usepackage{float}
\usepackage{amsmath}
\usepackage{graphicx}

\usepackage{babel}

\begin{document}

\title{Using the wavelet transform to separate scales in the Schrödinger equation and subsequently derive the Boltzmann equation}

\author{A.\,P. Meilakhs} 
\email[A.\,P. Meilakhs: ]{iestfi@gmail.com}
\affiliation{$^1$Ioffe Institute, 26 Politekhnicheskaya, St. Petersburg 194021, Russian Federation}
\date{\today}

\begin{abstract}
Here we continue our investigation aimed at the derivation of the Boltzmann equation from the first principles of quantum mechanics. The key difficulty that we overcome in this manuscript is the incompatibility between traditional methods employed in the derivation of classical mechanics from quantum mechanics and methods used to describe transitions between quantum states due to interaction. Here we use a novel technique that is similar to the wavelet transform method that originally appeared in the context of signal processing. With this method, we can simultaneously work with slowly varying potentials that give rise to classical-like wave-function behaviour, motion along a prescribed trajectory, and sharp potentials that lead to behaviour that is best understood in terms of quantum transitions. Working with smooth potentials in this framework, we derive the Liouville equation from the Schrödinger equation. When using this method for sharp potentials, we naturally obtain the description of system dynamics in terms of a transition-rate matrix.  The transition rates between system states, which characterize the process, are determined by Fermi's golden rule. We observe that, through the Liouville equation, we can deduce the non-collision part of the Boltzmann equation, and that, through the transition-rate matrix, we can deduce the collision integral. We combine the results to derive the  Boltzmann equation, the main formula of physical kinetics. 
\end{abstract}

\maketitle

\section{Introduction}

The arrow-of-time problem, as it is stated in common statistical mechanics and thermodynamics textbooks \cite{LandauV5, KondepudiPrigogine} as well as in specialized books on the nature of time \cite{Zeh1999, Kuzemsky2022}, can be formulated as follows: how is it possible that macroscopic processes are irreversible, while microscopic processes are time-symmetric? In our previous work, in the context of heat transport through an interface, we connected the irreversibility of this process with the concept of optical non-coherence \cite{Me24transmission, Me25Larotonda}. Subsequently, we extended these concepts from the interfacial case to the case of a bulk medium \cite{Me24NCevo}. This allowed an explanation of the irreversibility of transport processes in bulk.

Most notably, following this route, we derived the Boltzmann collision integral in particular cases in \cite{Me24NCevo}. However, we only investigated the case in which the considered system is spatially homogeneous. This means that there are no external fields, no concentration gradients, and no other violations of homogeneity. In this manuscript, we aim to remove this limitation. Our goal is to derive the full Boltzmann equation, which contains the collision, force, and diffusion terms, starting from the Schrödinger equation.

There are two obvious obstructions to reaching this goal. Firstly, the distribution function, whose evolution is described by the Boltzmann equation, is a function of coordinates and momenta. Meanwhile, the psi-function in the Schrödinger equation only depends on coordinates, and momenta are represented by differential operators. Secondly, in the Boltzmann equation, there is a clear distinction between short-scale interactions that contribute to the collision term and the external field that contributes to the force term. However, there is just a single potential energy term in the Hamiltonian.

We are going to overcome these problems with the help of the wavelet transform \cite{Mallat2008, Daubechies1992, Meyer1992}. 
The continuous wavelet transform consists of the integration of the product of a given function and a wavelet function. It is thus a variation of Fourier methods, but with distinct features. The wavelet function is non-zero only in a finite interval (or equivalently, the integration is carried out in a finite interval) and depends on two parameters (in one-dimensional space): the starting point of the interval and the frequency parameter. The integration is carried out for all possible values of both parameters; thus, if we initially had a function of one variable, the wavelet transform of it will be a function of two parameters.

This property of the wavelet transform turns a one-variable psi-function into a two-variable distribution function, and with just a little work we will show how the differential operator transforms into classical momentum. Concerning the second obstruction, we find that when the characteristic scale of interaction is sufficiently larger than the length of the integration interval, there arise some specific simplifications. And in the opposite case, when the characteristic scale of interaction is sufficiently shorter than the length of the integration interval, there arise completely different simplifications. Thus, short-scale and long-scale interactions in the Schrödinger equation behave qualitatively differently. This is why they contribute to different terms in the Boltzmann equation. We call this "the separation of scales" and consider it the most important result of the manuscript.

Wavelet theory emerged from problems in signal processing and became a highly popular area of research in the 1980s \cite{GrossmannMorlet1984, Daubechies1988, Mallat1989}. It remains an active area of research, with ongoing developments in multiscale representations, adaptive wavelet constructions, and computationally efficient algorithms \cite{Arts2022, Raath2023, LucasGilles2024, GillesTranOsher2024, Nguyen2024}. It is also an indispensable tool in a variety of applications, especially in biomedicine \cite{Shuvo2025application, Agyemang2025, Qin2025, Hou2025, Zhang2025seismic}. Since the goals of our investigation differ substantially from the typical goals of signal processing, there are also significant distinctions between the method that we use and traditional wavelet transforms. Most notably, in signal processing, the length of the wavelet function shrinks with the growth of the frequency parameter. This allows for higher spatial resolution for the wavelet transform at high frequencies. We, however, keep the integration interval constant so the separation of scales does not depend on frequency. Otherwise, we consider many-particle psi functions that possess symmetric or antisymmetric properties depending on the particles being fermions or bosons. This requires a specific technique that is absent in standard signal processing applications.

We now want to show how the results of this manuscript are connected to other modern research directions in physics. We can identify two such connections. The first is the derivation of classical mechanics through the concepts of decoherence theory \cite{Joos2003appearance}. The problem at hand is the following: a wave packet does not fully represent the behavior of classical particles, since it can evolve into a superposition of states with different probabilities, something that does not occur for classical particles. Decoherence theory attempts to overcome this problem by introducing interactions with the environment as a mechanism for the collapse of the wave function into one particular state \cite{Zurek1993}. Several different approaches exist \cite{Zurek1981packet, Joos1985packet, Omnes2011packet, Zurek2013packet}, along with a well-known set of objections \cite{Pessoa1997contr, Adler2003contr, Wallace2012contr}. The major objection is that decoherence explains only the vanishing of the off-diagonal elements of the density matrix, but does not explain why a single definite outcome is realized in a measurement.

The goal of the theory we present is to deduce the Boltzmann equation from the Schrödinger equation. The Boltzmann equation describes the evolution of single-particle distribution functions, which represent the mean number of particles in certain regions of phase space. In our theory, the mean numbers of particles are obtained through averaging over the probability distribution of particle positions given by the square of the absolute value of the $n$-particle psi-function. To derive the Boltzmann equation in such a manner, we only need to explain the classical-like behavior of each sub-wave-packet that represents a state; we do not need the exact outcome or a collapse of the state to a single wave packet. Hence, we do not encounter the outcome problem.

We note that the general research direction, which we outlined at the beginning of the section, overlaps significantly in goals and methods with one branch of decoherence theory, which considers quantum measurements to be the source of irreversibility \cite{Schulman1997, joshi2022probing}. One can also find similarities with quantum thermodynamics, which explains irreversibility as a result of the interaction of a system with its environment \cite{breuer2002openquant, rigol2008thermalization, deutsch2010thermodynamic, dymarsky2018subsystem}. However, in this particular manuscript, the overlap is not especially important. For an extensive discussion of the similarities and distinctions between our approach and the approaches of decoherence theory and quantum thermodynamics, see the first and last sections of \cite{Me24NCevo}.

The second line of research is the derivation of the Boltzmann equation from classical mechanics, namely from the Liouville equation. The classical approach involves the introduction of the BBGKY hierarchy \cite{Harris2004} and the application of Grad’s limit \cite{Grad1958}. The reasoning goes as follows: the Liouville equation describes the evolution of the full $N$-particle distribution function, the BBGKY hierarchy provides a way to deduce the $(k-1)$-particle distribution function from the $k$-particle distribution function under some reasonable assumptions, and the Boltzmann equation concerns the single-particle distribution function. What we do is take the last equation of the BBGKY hierarchy, namely the one that relates the single-particle distribution function to the two-particle distribution function. Then we replace the two-particle distribution function with the product of two single-particle distribution functions. This last part of the derivation, known as the \textit{Stosszahlansatz} or molecular chaos hypothesis, also introduces the direction of time.

Consequently, the possibility of representing a $k$-particle distribution function as a product of single-particle distribution functions (the factorization assumption) attracted considerable research attention. The first rigorous result was presented by Lanford \cite{Lanford1975}, who proved, roughly speaking, that if an $N$-particle distribution function is given by a product of single-particle distribution functions at the initial time, and if the evolution of the $N$-particle distribution function is described by the Liouville equation while that of the single-particle distribution function is described by the Boltzmann equation, then after a short period of time the $N$-particle distribution function is still close to a product of single-particle distribution functions. The derivation has been refined several times \cite{IllnerPulvirenti1989Refine, Marklof2011Refine, Mischler2013Refine, Bodineau2017Refine, Gallagher2014RefineBook} with different sets of simplifying assumptions, such as the proximity of the gas to vacuum or to the equilibrium state. This research program recently culminated in the manuscripts \cite{Deng2024Hil6, Deng2025Hil6}, which presented a proof valid for an arbitrarily long time interval without additional simplifications.

Unfortunately, this does not seem to resolve the contradiction between the time symmetry of classical mechanics and the time asymmetry of the Boltzmann equation. At least since \cite{Cohen1960InCondAsymm}, it has been known that by slightly modifying the derivation of the Boltzmann equation one can change the resulting sign of the collision integral and thus choose a different direction of entropy growth in time. Both approaches, however, rely equally on the factorization assumption. As Lanford himself put it, "...we can also study the limiting behavior for negative times. Exactly the same analysis applies as for positive times except that, [\ldots] collision occurs immediately if the extra particle is inserted in an incoming collision configuration with the $i$-th particle and not if it is in an outgoing configuration. This has the effect of changing the sign of the limiting collision term" and then "irreversibility appears [\ldots] not in the assumption that the limiting rescaled correlation functions factorize" (the end of Section 7 in \cite{Lanford1975}).

In our theory, we resolve the contradiction between the reversibility of classical mechanics and the irreversibility of the Boltzmann equation in a completely different way. In our previous paper \cite{Me24NCevo}, we proposed a derivation of the collision-integral part of the Boltzmann equation from quantum mechanics. We specifically addressed the problem of the origin of irreversibility: it arises from the non-coherence of interacting psi-functions. And the non-coherence, in this treatment, is a consequence of a finite spectral width of quantum states. However, we note that this theory is applicable only to interactions with sufficiently high-frequency fields, or, otherwise, to short-scale interactions.

In the present paper, we describe the interaction of the system with sufficiently low-frequency fields. This leads to the derivation of the Liouville equation from the Schrödinger equation. When we combine both scales into a single equation, we find that the long-wavelength potentials lead to the force term, and the short-wavelength potentials correspond to the collision part of the Boltzmann equation. Of these two parts, it is the collision one that is responsible for irreversibility. Classical mechanics, in the form of the Liouville equation, is in the other, long-scale limit. Thus, irreversible behaviour and classical-like behaviour arise in different limits of the Schrödinger equation and lead to different parts of the Boltzmann equation, and there is no contradiction.

In the subsequent two sections, we need to revisit the seemingly old and complete problem of deriving classical mechanics from quantum mechanics. The one-dimensional case is treated in Section II, and the generalization to an arbitrary number of dimensions is given in Section III. In Section IV of the manuscript, we deal with short-scale interactions. In Section V, we combine all the results obtained here with our earlier results to derive the full Boltzmann equation from quantum mechanics.

\section{Long wavelength limit, one-dimensional single-particle case}

We consider the one-dimensional Schrödinger equation for a single particle with slowly-varying potential. We seek the spatial part of the wave function in the form
\begin{equation}
\Psi (x) = \sum_p A_p(x) \phi_p(x).
\label{Psi}
\end{equation}
 Here $A_p(x)$ is a smooth envelope of the wave function that determines the probability amplitude for a particle to have coordinate and momentum around $x, p$, and $\phi_p(x) = \exp(\frac{i}{\hbar}(px-Et))$, where $E = p^2/2m + U$, is the fast oscillating part.

We note that a similar form of the wave function is used in the derivation of the Fermi rule, which describes the rate of quantum transitions. The difference is that the latter aims to separate the slowly varying and fast oscillating parts in time, so one studies $A_E(t)$, while here we are interested in $A_p(x)$. Deriving classical mechanics using the same language that is used to describe quantum transitions is the key result of the manuscript.

We take into account that a general wave function need not be an eigenfunction of the Hamiltonian and therefore have a specified value of energy. Instead, we assume the energy belongs to a small interval of values. And that is why the psi-function should not have one specific value of $p$ at a given point; hence we write the summation over $p$ in our ansatz. This is completely analogous to writing the psi-function as a wave packet, with momenta belonging to some interval and energies belonging to some interval.

We start with the Schrödinger equation in the absence of an external field, that is, for $U(x) = 0$. We substitute the wave function in the form \eqref{Psi}. For the time derivative of the $\Psi$-function, we have
\begin{equation}
i \hbar \frac{\partial}{\partial t} \Psi = i\hbar \dot A \phi_p(x) + E\Psi,
\label{time}
\end{equation}
For space derivative we have
\begin{equation}
- \frac{\hbar^2}{2m} \frac{\partial^2}{\partial x^2 }\Psi = 
\left( - \frac{\hbar^2}{2m} A''  - i\hbar \frac{p}{m} A' + \frac{p^2}{2m} A \right) \phi_p(x).
\label{space}
\end{equation}
We omitted the summation over $p$, since the equations are correct for every value of $p$ separately.

We combine equations (\ref{time}, \ref{space}). The third term in brackets in Eq. \eqref{space} cancels out with $E\Psi$. Since $A$ changes slowly with coordinate, the first term in Eq. \eqref{space} is much smaller than the second term. Taking it into account, we obtain
\begin{equation}
\dot A = - \frac{p}{m} A'
\label{preliminary}
\end{equation}

We have a similar equation for $A^*$, the complex conjugate of $A$. We multiply Eq. \eqref{preliminary} by $A^*$, and we multiply the equation for $A^*$ by $A$, and add both equations. We get
\begin{equation}
\dot \rho = - \frac{p}{m}\rho'
\label{preliminary2}
\end{equation}
where $\rho$ is the probability density of a particle to have coordinate and momentum around $x, p$, it is expressed as $\rho = A^*A$.

We now want to add the potential term. At this point, we need to introduce the concept of "wave packets" into the theory. We take into account that, in general, near some given point $x_0$, the $\Psi$-function may contain many components with different values of $p$, as in equation \eqref{Psi}. We substitute the $\Psi$-function in this form into the Schrödinger equation, and we do not write the kinetic-energy term, since we took care of it earlier:
\begin{equation}
i \hbar \frac{\partial}{\partial t} \sum_p A_p(x) \phi_p(x) = U(x) \sum_p A_p(x)\phi_p(x) + \dots
\label{Shred}
\end{equation}

We want to “cut out a piece’’ of the wave function in a coordinate region near $x_0$ and a momentum region near $p_0$, so we will use a version of the wavelet transform. We multiply the equation by $\phi_{p_0}^*(x)$ and integrate it over the region from $x_0$ to $x_0 + \Delta x$. We will then introduce some simplifications, which are applicable if a scale $\Delta x$ is large compared to the wavelengths of the fast-oscillating part of the $\Psi$-function, but small compared to the characteristic scale of variation of the amplitude.

 We notice that 
\begin{align}
\int \limits_{x_0}^{x_0 + \Delta x} \phi_{p_0}^*(x) A_p(x) \phi_p(x) \approx 
A_p(x_0) \int \limits_{x_0}^{x_0 + \Delta x} \phi_{p_0}^*(x) \phi_p(x) = 
- \frac{i\hbar}{p - p_0}\left(e^{\frac{i}{\hbar}(p - p_0) \Delta x} - 1 \right) A_p(x_0).
\label{smoothing}
\end{align}
The expression inside the brackets, as a function of $p$, is large in the vicinity of $p_0$ and small away from $p_0$. If we consider $A_p$ to be a slowly varying function of $p$, the value in the vicinity of $p_0$ can be approximated as a constant. Thus, we can put
\begin{equation}
\int_{x_0}^{x_0 + \Delta x} \phi_{p_0}^*(x) \sum_p A_p(x)\phi_p(x) \approx  \int dp \delta(p - p_0) A(x, p) = A(x_0, p_0).
\label{delta}
\end{equation}
Here we assumed that $\Delta x$ is large compared to the scale of the fast-varying part of the $\psi$-function $\phi_p(x)$, so the integral over the range $\Delta x$ behaves approximately as an integral over an infinite range. Here $A(x_0, p_0)$ is now a smooth function of two variables.

That is our expression for the left-hand side of the equation. For the right-hand side, we consider the potential $U(x)$ to be a smooth function, so we expand it into a Taylor series near $x_0$ and keep only the first term
\begin{equation}
\int_{x_0}^{x_0 + \Delta x}  \phi_{p_0}^*(x) A_p(x) U(x)\phi_p(x) = \int_{x_0}^{x_0 + \Delta x}  \phi_{p_0}^*(x) A_p(x)\left(U(x_0) + \frac{\partial U}{\partial x} x \right) \phi_p(x).
\label{Taylor}
\end{equation}
The part with $U(x_0)$ is cancelled with the energy term on the left side: $E = p^2/2 + U(x_0)$. For the term with derivative with respect to $x$, after we have taken $\frac{\partial U}{\partial x}$ at $x_0$ outside of the integration limits, we have
\begin{align}
 \int_{x_0}^{x_0 + \Delta x}dx \phi_{p_0}^*(x) \left( \sum_p  A_p(x) x \phi_p(x) \right) 
= \int_{x_0}^{x_0 + \Delta x}dx \phi_{p_0}^*(x) \left( \sum_p A_p(x) \left[  -i \hbar \frac{\partial}{\partial p}\phi_p(x) \right] \right) = \nonumber \\
 -i \hbar \sum_p  A_p(x) \frac{\partial}{\partial p}  \left(\int_{x_0}^{x_0 + \Delta x} dx  \phi_{p_0}^*(x)   \phi_p(x) \right) =  -i \hbar \Delta x \int_p \frac{dp}{2\pi\hbar}  \frac{\partial}{\partial p} \left( \frac{2\pi\hbar}{ \Delta x}\delta(p - p_0) \right) A(x, p) = i \hbar \frac{\partial}{\partial p_0}A(x_0, p_0).
\label{derivative}
\end{align}
Here, the multiplier $\Delta x/(2\pi\hbar)$ appears because of the transition from summation to integration. This derivation relies on the interchangeability of differentiation and taking the weak limit (delta function) \cite{Zemanian2011}.

Instead of taking the derivative of a delta function, we can use a longer and less rigorous, yet more demonstrative and intuitive approach. For this, we employ a technique developed in \cite{Me24NCevo}. We divide the expression preceding $A$ on the right side of equation \eqref{smoothing} by $\Delta x$. We write
\begin{align}
- \frac{i\hbar}{\Delta x(p - p_0)}\left(e^{\frac{i}{\hbar}(p - p_0) \Delta x} - 1 \right) = \chi_{\Delta p}(p_0),
\end{align}
and observe that
\begin{align}
\Delta p \Delta x \gg \hbar =>\chi_{\Delta p}(p_0) \approx 0, \nonumber \\
\Delta p \Delta x \ll \hbar => \chi_{\Delta p}(p_0) \approx 1.
\label{cases1}
\end{align}
Now to simplify the expression, instead of $\chi_{\Delta p}(p_0)$ we introduce an approximation $\overline \chi_{\Delta p}(p_0)$ which is an indicator function of a segment of momenta in the vicinity of $p_0$
\begin{align}
\Delta p \Delta x > &\pi\hbar  => \overline \chi_{\Delta p}(p_0) = 0 \nonumber \\
\Delta p \Delta x < &\pi\hbar  => \overline \chi_{\Delta p}(p_0) = 1.
\label{cases2}
\end{align}
It is a simultaneous definition of the indicator function in momentum space and of the width of the step $\Delta p$. This exchange is possible because both $\chi_{\Delta p}$ and $\overline\chi_{\Delta p}$ form a delta sequence and weakly converge to the delta function in the limit $\Delta p \to 0$.

For the first part of the derivation \eqref{delta}, this does not introduce anything new. We substitute it into the second expression from \eqref{derivative} and obtain
\begin{align}
 \int_{x_0}^{x_0 + \Delta x} \phi_{p_0}^*(x) \sum_p A_p(x) x \phi_p(x) =
 -i \hbar  \sum_p \frac{\partial}{\partial p} \left(\int_{x_0}^{x_0 + \Delta x} \phi_{p_0}^*(x) \phi_p(x) \right) A_p(x) =
 -i \hbar  \sum_p \Delta x \frac{\partial}{\partial p} \overline \chi_{\Delta p}(p_0) A_p(x) =  \nonumber \\ 
=  -i \hbar \int_p dp \frac{1}{2 \Delta p}\left( -\delta (p_0+\Delta p) + \delta (p_0 - \Delta p) \right) A(p, x_0) =  
 i \hbar \frac{A(p_0+\Delta p) - A(p_0 - \Delta p)}{2 \Delta p} =  i \hbar \frac{\partial}{\partial p_0}A(x_0, p_0),
\label{derivative2}
\end{align}
where we have used the formula for the derivative of a step function, and the last equality is the definition of the smoothed derivative. Thus, we arrive at the same conclusion in two slightly different ways.

Let us reflect on what we have achieved so far. We started with the $\psi$-function, which was a function of one variable, the coordinate, while momentum was represented as an operator. In the presence of external fields, momentum $p$ was an index labeling the states of the system. We partitioned the $\psi$-function into two parts -- the fast-oscillating part and a smooth envelope -- and derived an equation for the latter. It turns out that it is a function of two variables, coordinate and momentum. When we consider the equation for the smooth envelope, momentum is no longer an operator but a variable, as in classical mechanics!

We proceed to obtain $\rho$ instead of $A$ in the same manner as before, and combine it with Eq. \eqref{preliminary} to get
\begin{equation}
\dot \rho = - \frac{p}{m} \frac{\partial \rho}{\partial x} + \frac{\partial U}{\partial x} \frac{\partial \rho }{\partial p}
\label{final}
\end{equation}

If we introduce a Hamiltonian function (not operator!) $H = p^2/2m + U(x)$, we obtain
\begin{equation}
\dot \rho + \frac{\partial H}{\partial p} \frac{\partial \rho }{\partial x} - \frac{\partial H}{\partial x} \frac{\partial \rho }{\partial p} = 0,
\label{1DLiuv}
\end{equation}
the one-dimensional Liouville  equation.

We can derive Hamiltonian mechanics from this point, as a somewhat inverse form of Liouville's theorem. We use the method of characteristics \cite{ArnoldODE} to get a system of ordinary differential equations equivalent to this partial differential equation. We get
\begin{align}
&\frac{d x}{d s} = \frac{\partial H}{\partial p} \nonumber \\
&\frac{d p}{d s} = -\frac{\partial H}{\partial x}  \nonumber \\
&\frac{dt}{d s} = 1  \nonumber \\
&\frac{d \rho}{d s} = 0.
\label{Hamilton}
\end{align}
From the third equation we obtain $dt = ds$, so the first two equations reduce to Hamilton’s equations, while the fourth equation becomes the equation expressing conservation of probability.

While this result is trivial, especially when Liouville’s theorem is formulated in the mathematical language of Hamiltonian vector fields \cite{ArnoldMech}, we nevertheless include this derivation. It seems worthwhile, since in the physical literature authors are typically interested in deducing Liouville’s equation from Hamilton’s equations, and not the other way around.

The essence of the method of characteristics is the transition from a field description, which assigns to each point in space–time the value of the field, to a particle description, which assigns to every particle a trajectory in space–time. If we substitute the delta function in coordinate and momentum into the Liouville equation, it will continue to propagate as a delta function along the characteristic curve, and the equation of this characteristic curve is Hamilton’s equation.

In the standard derivation of classical mechanics from quantum mechanics, based on wave packets, delta-function motion along the classical trajectory corresponds to the motion of the center of coordinates and momenta of a sufficiently localized wave packet. However, Liouville’s equation admits arbitrary initial conditions, not just a delta function. In this sense, the presented theory generalizes the standard wave-packet description.

We also note that the wave-packet description includes spreading of wave packets. In our formalism, the term with the second derivative $- \frac{\hbar^2}{2m} A''$ in Eq. \eqref{space}, that we have excluded as small, corresponds to this phenomenon.

\section{Full Liouville equation}

Now we want to consider the many-particle case. Let us consider the case of fermionic particles, so the fast oscillating part of the  $\psi$-function can be expressed as a Slater determinant:
\begin{equation}
\psi_\mathbf{p}(\mathbf{x}) = \frac{1}{n!} \mathrm{Det} \begin{pmatrix} \phi_1(x_1) & \phi_1(x_2) & ... & \phi_1(x_n) \\ 
\phi_2(x_1) & \phi_2(x_2) & ... & \phi_2(x_n) \\
... & ... & ... & ...\\
\phi_n(x_1) & \phi_n(x_2) & ... & \phi_n(x_n) \end{pmatrix},
\label{determinant}
\end{equation} 
where  $\mathbf{x}$ stands for all the set $x_1,.. x_n$ and $\mathbf{p}$ for all the set $p_1,.. p_n$ and $\phi_k(x_l) = \exp(\frac{i}{\hbar}p_k x_l)$.

 The full wave function that contain the fast oscillating part $\psi_ \mathbf{p}(\mathbf{x})$ and the smooth envelope $A(\mathbf{x},  \mathbf{p})$ is expressed as
\begin{equation}
\Psi (\mathbf{x}) = \sum_\mathbf{p} A(\mathbf{x}, \mathbf{p}) \psi_\mathbf{p}(\mathbf{x}),
\label{PsiFull}
\end{equation}

The kinetic energy term can be expressed as 
\begin{equation}
- \frac{\hbar^2}{2m} \left( \frac{\partial^2}{\partial x_1^2 } + \frac{\partial^2}{\partial x_2^2 } + 
... + \frac{\partial^2}{\partial x_n^2} \right).
\label{KinOp}
\end{equation}

When we apply the kinetic energy operator \eqref{KinOp} to the $\psi$-function in the form (\ref{determinant}, \ref{PsiFull}), we obtain three types of summands, as in expression \eqref{space}. In the terms of the first type, the derivatives are applied twice to the slowly varying amplitude $A$, so the result is negligible. If the derivative is applied twice to the $\psi$-function \eqref{determinant}, we obtain a contribution to the total kinetic energy of the system. The sum of such contributions is the total kinetic energy, and it cancels out with the time derivative of $\psi$, as in the one-particle case (\ref{time}, \ref{space}). Thus, the only terms that we should take care of are those in which one derivative is applied to $A$ and one is applied to $\psi$. The terms of this type have the form
\begin{equation}
\frac{i \hbar}{m} \frac{\partial A}{\partial x_l}
  \frac{1}{n!} \mathrm{Det} \begin{pmatrix} \phi_1(x_1)  & ... & p_1 \phi_1(x_l) & ... & \phi_1(x_n) \\ 
 \phi_2(x_1)  & ... & p_2 \phi_2(x_l) & ... & \phi_2(x_n) \\
... & ... & ... & ... & ... \\
 \phi_n(x_1)  & ... & p_n \phi_n(x_l) & ... & \phi_n(x_n) \end{pmatrix}.
\label{Summand}
\end{equation} 
Since the full fermionic psi function \eqref{PsiFull} is antisymmetric with respect to the interchange of the coordinates and the oscillating part is antisymmetric (as a property of the determinant), the amplitude should be symmetric. Hence it follows
\begin{equation}
\frac{\partial A}{\partial x_l} =  \frac{\partial A}{\partial x_k}.
\label{Symmetry}
\end{equation}
for all pairs of indices $k, l$. 

We expand the total sum of all terms of the form \eqref{Summand}, and we expand each determinant along the column with momenta. We denote by $M_{ij}$ the determinant of the minor obtained from \eqref{determinant} after erasing the $i$-th row and the $j$-th column. We use the Laplace rule for the signs in the expansion and obtain:
\begin{align}
& \frac{\partial A}{\partial x_1} \left(p_1 \phi_1(x_1) M_{11} - p_2 \phi_2(x_1) M_{21} + ... + (-1)^{n+1} p_n \phi_n(x_1) M_{n1} \right) + \nonumber \\
+ & \frac{\partial A}{\partial x_2} \left(- p_1 \phi_1(x_2) M_{12} + p_2 \phi_2(x_2) M_{22} - ... + (-1)^{n+2} p_n \phi_n(x_2) M_{n2} \right) + ...
\end{align} 
We expand the brackets and rearrange the partial derivatives with the help of expression \eqref{Symmetry} in such a way that the index of the derivative coincides with the index of the momentum:
\begin{align}
& \frac{\partial A}{\partial x_1} p_1 \phi_1(x_1) M_{11} - \frac{\partial A}{\partial x_2} p_2 \phi_2(x_1) M_{21} + ...  \nonumber \\
- & \frac{\partial A}{\partial x_1} p_1 \phi_1(x_2) M_{12} + \frac{\partial A}{\partial x_2} p_2 \phi_2(x_2) M_{22} - ... 
\end{align} 
We finally group members with the same index of momentum and derivative
\begin{align}
& \frac{\partial A}{\partial x_1}p_1 \left( \phi_1(x_1) M_{11} - \phi_1(x_2) M_{12} + ... + (-1)^{n+1}\phi_1(x_n) M_{1n} \right) + \nonumber \\
& \frac{\partial A}{\partial x_2}p_2 \left( - \phi_2(x_1) M_{21} + \phi_2(x_2) M_{22} + ... + (-1)^{n+2}\phi_2(x_n) M_{2n} \right) + ...
\end{align} 
The expressions in the brackets are nothing but the expansions of the determinant along the rows. The expression in the first line of the equation is the expansion along the first row, the expansion in the second line is the expansion along the second row, and so on. However, every such expansion is just the same determinant $\psi_ \mathbf{p}(\mathbf{x})$ \eqref{determinant}. Thus, we have
\begin{align}
 \left(\frac{\partial A}{\partial x_1}p_1 + \frac{\partial A}{\partial x_2}p_2 + ... \right) \psi_ \mathbf{p}(\mathbf{x}).
\end{align} 
We note that the same expression would arise for bosons, except that in this case the determinant would be replaced by the permanent in the expression for $\psi_\mathbf{p}(\mathbf{x})$ \eqref{determinant}. The permanent is analogous to the determinant, but without the alternating signs between summands. All that we need to make an analogous derivation is the ability to expand the permanent by minors along a row or a column, a property that it shares with the determinant.

Since the total psi-function should be symmetric and the permanent is also symmetric, the amplitude should be symmetric as well, so that expression \eqref{Symmetry} holds. Otherwise, the derivation is the same. We note that in books on physical kinetics \cite{Harris2004}, the distribution function in Liouville's equation is assumed to be symmetric. Here, we have justified this assumption, both for fermions and bosons.

For the momentum part of our derivation, we use the many-particle wavelet transform. We introduce a region of space bounded by $x_{01}$ and $x_{01} + \Delta x_{01}$, $x_{02}$ and $x_{02} + \Delta x_{02}$, and so forth. Furthermore, when we use integration over coordinates, we mean integration over this region. We expand the potential function inside this region as
\begin{equation}
U(\mathbf{x}) = U(\mathbf{x}_0) + \frac{\partial U}{\partial x_1} x_1 + ... + \frac{\partial U}{\partial x_n} x_n.
\label{TaylorND}
\end{equation}
We notice, that 
\begin{equation}
x_l \psi_\mathbf{p}(\mathbf{x}) =
 \mathrm{Det} \begin{pmatrix}
 \phi_1(x_1)  & ... & \frac{\partial}{\partial p_1} \phi_1(x_l) & ... & \phi_1(x_n) \\ 
 \phi_2(x_1)  & ... & \frac{\partial}{\partial p_2} \phi_2(x_l) & ... & \phi_2(x_n) \\
... & ... & ... & ... & ... \\
 \phi_n(x_1)  & ... & \frac{\partial}{\partial p_n} \phi_n(x_l) & ... & \phi_n(x_n) \end{pmatrix} =
\sum_k \frac{\partial}{\partial p_k} \phi_k (x_l) M_{kl}.
\label{MinorsX}
\end{equation} 
Also we notice that
\begin{equation}
 \int d x_l \phi_m^* (x_l) \frac{\partial}{\partial p_k}  \phi_k (x_l) = \delta_{mk} \delta(p_k - p_{0k}) \frac{\partial}{\partial p_k} .
\label{IntDeltaN}
\end{equation}
Here, the derivative with respect to $p_k$ acts on $A(p_k)$ as in formula \eqref{derivative}.

We multiply the Schrödinger equation over $\psi_\mathbf{p_0}(\mathbf{x})^\dag$, integrate over a region of space, substitute the multidimensional Taylor expansion for potential and, combining everything above, get:
\begin{align}
 \int d\mathbf{x} \psi_\mathbf{p_0}^\dag(\mathbf{x}) \left ( \sum_\mathbf{p} A_\mathbf{p}(\mathbf{x}) \left[ \sum_i \frac{\partial U}{\partial x_i} x_i \right] \psi_\mathbf{p}(\mathbf{x})  \right)
 =
\int d\mathbf{x} \psi_\mathbf{p_0}^\dag(\mathbf{x}) \left ( \sum_\mathbf{p} A_\mathbf{p}(\mathbf{x}) \left[ \sum_i \frac{\partial U}{\partial x_i} \sum_j \frac{\partial }{\partial p_j}  \phi_j  M_{ij} (\mathbf{x}) \right] \right)
=
 \nonumber \\ 
=    \sum_\mathbf{p} A_\mathbf{p}(\mathbf{x})  \sum_i \frac{\partial U}{\partial x_i} \sum_j  \left[ \int d\mathbf{x} \psi_\mathbf{p_0}^\dag(\mathbf{x}) \frac{\partial }{\partial p_j}  \phi_j  M_{ij} (\mathbf{x}) \right] 
=
  \sum_\mathbf{p} A_\mathbf{p}(\mathbf{x})  \sum_i \frac{\partial U}{\partial x_i} \prod_{j\neq k} \delta(p_j - p_{0j})  \delta_{ik}  \frac{\partial}{\partial p_k} \delta(p_k - p_{0k})  =
 \nonumber \\ 
= \sum_i \frac{\partial U}{\partial x_i} \frac{\partial}{\partial p_i} A_{\mathbf{p}_0}(\mathbf{x}_0)  
\end{align}

Overall, we get the equation of evolution of the amplitude for arbitrary many degrees of freedom:
\begin{equation}
\dot A + \sum_{i=1}^n \left( \frac{\partial H}{\partial p_i} \frac{\partial A }{\partial x_i} - \frac{\partial H}{\partial x_i} \frac{\partial A }{\partial p_i} \right) = 0.
\label{AmpLiouville}
\end{equation}
We see that this equation can be turned into the Liouville equation for the probability density, with the same manipulation we used before in Eqs. (\ref{preliminary}, \ref{preliminary2}).

Before moving to the next part, let us review the development so far. When deriving classical mechanics from quantum mechanics, standard textbooks on the subject \cite{QMBlokhintsev, QMDavydov} present one of three complementary types of reasoning. One is the quasiclassical approximation, which consists of substituting the wave function in the form $\psi = \exp(iS/\hbar)$ into the Schrödinger equation and expanding it in a power series in $\hbar$. The zeroth term is the only one that survives in the limit $\hbar \to 0$, and it corresponds to the Hamilton–Jacobi equation of classical mechanics. Another is the Ehrenfest theorem, which provides relations between the coordinate and momentum operators analogous to the Hamiltonian equations of classical mechanics.

The last method considers a wave packet and the properties of its motion. It is shown that, under sufficiently good conditions, the probability density moves with velocity $\vec p/m$, and its acceleration is proportional to the gradient of the potential, just as in classical mechanics.

These methods, while they serve the purpose of presenting the transition from quantum to classical, still lack a small element each. The classical Hamilton–Jacobi equation defines a trajectory along which a classical particle actually moves, and along which the change of coordinates occurs. By contrast, the quasiclassical wave function is constant in time, up to a phase factor, so no redistribution of probability takes place. In the same spirit, Ehrenfest’s theorem is too general: it applies to any state of the system, even to stationary states of discrete spectrum, which look nothing like motion along trajectories in classical mechanics.

The wave packet is the concept that most closely resembles motion along a trajectory. However, it lacks generality. What happens when a wave packet encounters a potential barrier and is partly transmitted and partly reflected? If we assume that the potential is sufficiently smooth both before and after the barrier, it is intuitively clear that each part of the wave packet behaves like a classical particle. On the one hand, we can no longer call the state of the system classical, since it now consists of two parts, something that does not occur for a classical particle. On the other hand, the formal derivation of the proportionality between acceleration and the gradient of the potential uses a Taylor expansion of the potential in a small region in which the wave packet is localized. We cannot perform this expansion if the wave function is a superposition of several wave packets that are at a sufficient distance from each other.

Moreover, what if the system consists of many particles, and they are bosonic or fermionic, with the corresponding symmetry requirements for the wave function? Will this affect the behavior of the many-dimensional wave packet? What if the system has interactions that result in quantum transitions, so that probability is continuously redistributed between different wave packets?

Here, we presented a new derivation of classical mechanics from quantum mechanics. We provided an extension of the wave-packet theory that is able to address these questions. This derivation has the same generality as the quasiclassical derivation and also captures the actual motion along a trajectory, as in the wave-packet picture. Now we want to add quantum transitions, caused by interactions, to this picture.

\section{Short-distance interactions}

When dealing with slowly varying potentials, we used a Taylor expansion of the potential term in a region with accuracy up to the first order. This approximation is valid if the potential term changes gradually. The basic idea of analyzing short-distance interactions is the opposite: if all the interaction happens in a region much smaller than the region of integration, then there is no sufficient difference between the actual expression and the expression that we would have obtained if integrated over the whole space. And integration over the whole space is just a normal Fourier transform. So the wavelet transform gives us the Taylor method (with some additions) for slowly varying potentials and the Fourier method (with some additions) for sharp potentials. 

We will start with external potential, since the intuition described above, for this case works literally. The physical example may be the scattering of quasiparticles on uncharged lattice impurities in crystals. We will consider interaction between particles later, since it requires additional considerations. Analogous to slow-varying potentials, first we will consider the one-dimensional case.

Let the potential $U(x)$ be zero outside of a region of length $ d: U(x)=0$ if $x'<x<x'+d$. Let us choose the integration interval to be much greater than $ d$ -- $\delta \gg d$, and let it include the whole interval at which $U(x)$ is nonzero. Doing the same calculations that led to the formula \eqref{Taylor}, we obtain the left-hand side of the following expression:
\begin{equation}
\int_{x_0}^{x_0 + \Delta x}  \phi_{p_0}^*(x) \sum_p A_p(x) U(x)\phi_p(x) = \sum_p A_p(x) \int_{x_0}^{x_0 + \Delta x}  \phi_{p_0}^*(x) U(x) \phi_p(x) = \sum_q A_{p_0+q} V(q),
\label{Fourier}
\end{equation}
where $q=p-p_0$. The result is valid for all points except for the small interval of length $4d$, in which the potential intersects the integration interval but does not fully lie within it. Since $d/\Delta x \ll 1$ we can just ignore it.

\begin{figure}
\includegraphics[width=0.60\textwidth]{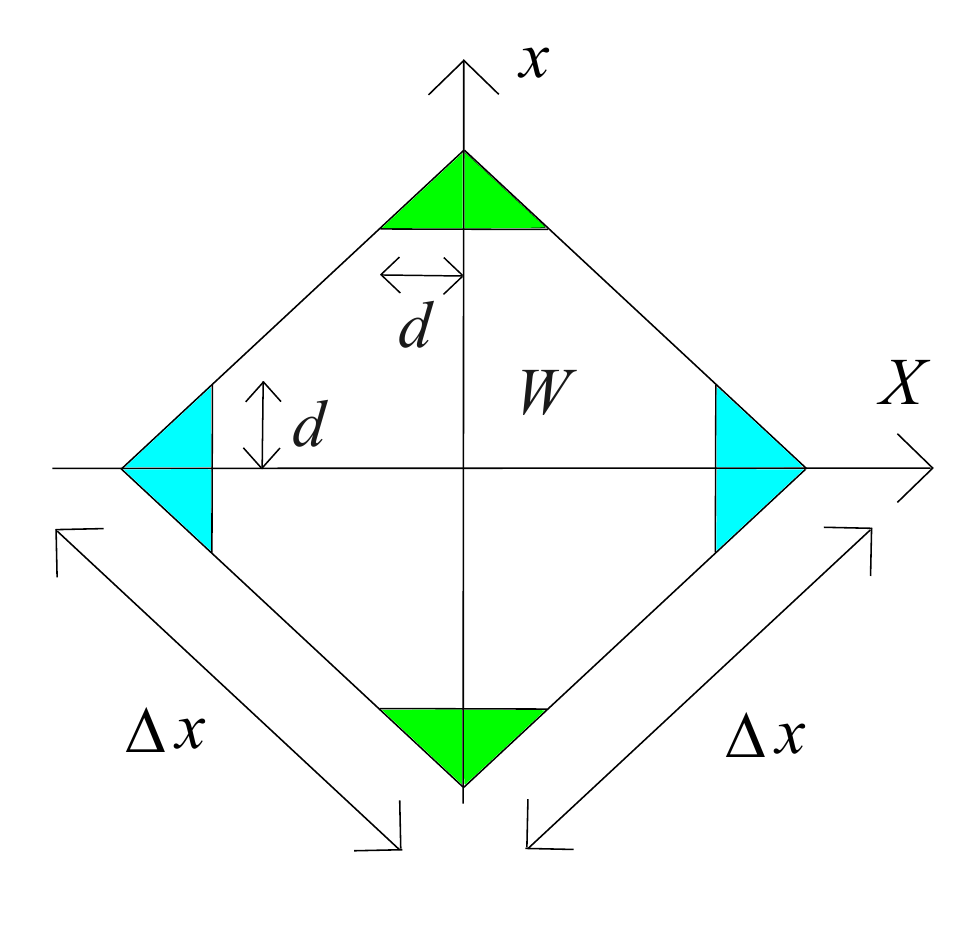}\caption{The integration region $W$ of Eq. \eqref{1DInteraction} is depicted. The blue regions at the limits of values of $X$ are those at which the range of integration over $x$ is small and thus the Fourier transform approximation fails. The green regions at the limits of values of $x$ are those at which the range of integration over $X$ is small and thus the conservation of momentum approximation fails. The characteristic scale for both types is $d$. It is evident that overall the error is of order $(d/\Delta x)^2$. }
\label{diagram}
\end{figure}

Now let us analyze a short-ranged interaction between two anyonic particles (with a non-symmetrized wave function) in one dimension. We have an interaction potential $U(x_1 - x_2)$ that equals zero when $|x_1 - x_2|>d$. We insert $\phi_p(x) = \exp(ipx)$, and use the coordinate change $X=x_1 + x_2$, $x = x_1 - x_2$. We denote $q = p_0 - p$. After multiplication and integration we obtain:

\begin{equation}
\int_{x_{10}}^{x_{10} + \Delta x} \int_{x_{20}}^{x_{20} + \Delta x} e^{iq_1x_1} e^{iq_2x_2} U(x_1 - x_2) = \iint_W dX dx e^{i(q_1 + q_2)X/2} e^{i(q_1 - q_2)x/2} U(x),
\label{1DInteraction}
\end{equation}
where $W$ denotes the region of integration. Normally, the integration over $X$ yields the conservation of momentum law $q_1 = -q_2 = q$, and integration over $x$ yields $V(q)$, the Fourier component of $U(x)$. So we would get the "normal result": $V(q)\delta(p_{01}+p_{02} - p_1-p_2)$. However, since the integration is not over the full plane, the result is different. In Figure \eqref{diagram} the region of integration is depicted. The deviation of the actual integral on the left-hand side of Eq. \eqref{1DInteraction} from the normal result arises when the range of integration over one of the variables is comparable to $d$. This happens at the corners of the region of integration.

For integration over $x$ we can apply the same logic as we did when analyzing expression \eqref{Fourier}. If the potential sits well within the boundaries of integration, specifically, at a distance greater than $d$ from the limit, we can extend the limit of integration over $x$ to infinity and obtain the Fourier component of the potential: $V(q)$. However, at high values of $X$, the limits of integration over $x$ shrink below $d$, so this approximation does not hold.

The integration over $X$, when carried out over a finite interval, is not equal to the delta function. Instead, as we have shown earlier (\ref{cases1}, \ref{cases2}), we can approximate it with the expression  $\overline \chi_{\Delta p} \dot \Delta X$. Here $\overline \chi_{\Delta p}$ is the indicator function of momentum with width $\Delta p$. $\Delta X$ is a region of integration over $X$ (with fixed value of $x$). The width of the indicator function is inversely proportional to the width $\Delta X$ of the integration region: $\Delta p = 2 \pi \hbar/ \Delta X$. If $V(q)$ varies slowly at the scale $\Delta p$, we can completely neglect the variation, which will give the same result as if we had just used a delta function. The width of $V(q)$ in momentum space is inversely proportional to the range of interaction $d$. So the criterion for using a delta function with very high precision is that the width of $V(q)$ is much greater than the width of $\Delta p$. Or, in the coordinates, it is $1/\Delta X \ll 1/d$. Or, if we do not need extremely high precision but just need the approximation to work, we can put it as $d \leq \Delta X$. The width of integration over $X$ is small when the value of $x$ is high. This is seen in Figure \eqref{diagram}. Overall, we conclude that the difference between the normal result and the actual expression is of order $(d/\Delta x)^2$.

Finally, we can deal with the full, unsimplified expression. Luckily, difficulties associated with the symmetry properties of the wave function do not overlap with difficulties associated with the finite integration region. A comprehensive derivation of the matrix element of a transition for a many-particle system is often omitted in textbooks, but can be found in Ref. \cite{NegeleOrland1988}. The standard expression for the matrix element of a transition for a many-particle system is 
$V(q) \delta(p_{01}+p_{02} - p_1- p_2) \hat a^\dag_{p_{01}} \hat a^\dag_{p_{02}} \hat a_{p_1} \hat a_{p_2} \prod_{i \neq 1,2} \delta(p_{0i} - p_i)$. \\ Here the $\hat a^\dag, \hat a$ denote the creation and annihilation operators, fermionic or bosonic, and the product over delta-functions means conservation of momentum for particles that are not involved in the interaction. Now, taking into account the finite integration limits, expression $V(q)$ is accurate with precision up to $(d/\Delta x)^2$, as we have shown earlier. The delta functions in the product should be replaced with $\overline \chi_{\Delta p_i} \Delta x_i$, meaning that conservation of $i-th$ momentum is not exact but to the precision of $2 \pi \hbar/ \Delta x_i $.

\section{Boltzmann equation}

We want to combine the results presented here with the results of the previous manuscript \cite{Me24NCevo} to derive the Boltzmann equation from quantum mechanics (two particular cases). However, we need to take care of the time-dependent part of the psi-function. We have
\begin{equation}
\Psi (\mathbf{x}) = \sum_\mathbf{p} A(\mathbf{x}, \mathbf{p}, t) \psi_\mathbf{p}(\mathbf{x}, t),
\label{PsiFinall}
\end{equation}
where $A$ is a slowly varying function of $t$ and $\psi$ is a fast-oscillating function. The dependence of $\psi$ on $t$ is given by the exponent
\begin{equation}
\psi_\mathbf{p} (t) = \exp \left(-\frac{iE_\mathbf{p}t}{\hbar} \right); E_\mathbf{p} = \sum_i \frac{p_i^2}{2m} + U(\mathbf{x}).
\label{Energy}
\end{equation}

Let the system in question have a smooth external potential and short-distance interactions between particles. Combining the results of the two previous Sections and multiplying both parts by $\exp \left(\frac{iE_\mathbf{p}t}{\hbar} \right)$, we obtain
\begin{equation}
\dot A_\mathbf{p} + \sum_{i=1}^n \left( \frac{p_i}{m} \frac{\partial A_\mathbf{p} }{\partial x_i} - \frac{\partial U}{\partial x_i} \frac{\partial A_\mathbf{p} }{\partial p_i} \right) = -\frac{i}{\hbar} \sum_\mathbf{p'} \exp\left(\frac{i}{\hbar}(E_\mathbf{p} - E_\mathbf{p'})t \right) \hat V_{\mathbf{p}\mathbf{p'}} A_\mathbf{p'}.
\label{AlmostThere}
\end{equation} 
Here $\hat V_{\mathbf{p}\mathbf{p'}}$ denotes the interaction operator that consists of a single-particle matrix element $V(q)$ and creation-annihilation operators, $\mathbf{p'}$ is assumed to differ from $\mathbf{p}$ in exactly four particle momenta, and conservation of momentum is already assumed in the summation. As a result of the separation of scales with the wavelet transform, the external potential and short-range interactions are contained in different terms of the expression. The external potential is on the left-hand side, while the short-range interactions are on the right-hand side.

We use the same wavelet transform method as previously, but with time instead of coordinates. We integrate expression \eqref{AlmostThere} over a time interval $t_{0} + \Delta t$ to get rid of the fast-varying part of the psi-function. This calculation was carried out thoroughly in Ref. \cite{Me24NCevo}. We arrive at
\begin{equation}
\dot A_\mathbf{p} +  \sum_{i=1}^n \left( \frac{p_i}{m} \frac{\partial A_\mathbf{p} }{\partial x_i} - \frac{\partial U}{\partial x_i} \frac{\partial A_\mathbf{p} }{\partial p_i} \right) = -\frac{i}{\hbar} \sum_\mathbf{p'} V_{\mathbf{p}\mathbf{p'}} A_\mathbf{p'}.
\label{Final}
\end{equation} 
where $V_{\mathbf{p}\mathbf{p'}}$ are many-particle matrix elements $ \langle \mathbf{p}|\hat V_{\mathbf{p}\mathbf{p'}}|\mathbf{p'} \rangle$ and the summation should be carried over states within  $\Delta E = \pi\hbar/\Delta t$ range of $E$. This expression concludes the use of the wavelet transform to deduce the equation for the slowly varying part of the psi-function.

Now we need to go from amplitudes to probability densities. For the left-hand side of Eq. \eqref{Final}, it is trivial, see Eqs.  (\ref{preliminary}, \ref{preliminary2}). However, for the right-hand side, we need an entirely different set of ideas. Namely, we use the concept of non-coherent evolution. This means that, at a given point, the probability amplitudes of different microstates are not phase-correlated with each other.

To explain this briefly, we recall that we are studying an extension of the wave-packet theory. Therefore, the state of the system consists of a sheaf of frequencies. When two states interact with each other, the interaction of each pair of single-frequency components is described, taking into account the phase difference between such components. However, the phase difference is a function of frequency. And for different frequencies, the phase difference is different. When calculating the mean value over all frequencies, if no phase correlation is present, we obtain values as if the states did not have amplitudes with phases, but only probabilities. The exact physical conditions that allow us to use this physical assumption, as well as its consequences, are described in \cite{Me24NCevo} in great detail.

With the non-coherence condition for the right-hand side, we arrive at:
\begin{equation}
\dot \rho_\mathbf{p} + \sum_{i=1}^n \left(  \frac{p_i}{m} \frac{\partial \rho_\mathbf{p} }{\partial x_i} - \frac{\partial U}{\partial x_i}  \frac{\partial \rho_\mathbf{p} }{\partial p_i} \right) = \sum_{\mathbf{p'}} Q_{\mathbf{p}\mathbf{p'}} \rho_\mathbf{p'}.
\label{FinalRhos}
\end{equation}
Here $Q_{\mathbf{p}\mathbf{p'}}$ are elements of the transition rate matrix. Transitions are between states of the system at given point and its elements are 
\begin{align}
Q_{\mathbf{p}\mathbf{p'}} = \frac{2 \pi}{\hbar} |V_{\mathbf{p}\mathbf{p'}}|^2 \delta(E_\mathbf{p} - E_\mathbf{p'}), \nonumber \\
Q_{\mathbf{p}\mathbf{p}} = -  \frac{2 \pi}{\hbar} \sum_l |V_{\mathbf{p}\mathbf{p'}}|^2 \delta(E_\mathbf{p} - E_\mathbf{p'}).
\label{Fermi}
\end{align}
For non-diagonal elements, these transition rates coincide with the Fermi rule, the diagonal values are such that the sum of all elements in each row equals zero. The left-hand side of Eq. \eqref{FinalRhos} is, obviously, the Liouville equation. Together, these parts constitute an evolution equation for the full N-particle distribution function.

The final goal is to deduce the equation for the single-particle distribution function, namely, the Boltzmann equation. Again, the non-collisional part follows from the Liouville equation trivially. The more interesting task is to deduce the collision integral form the right-hand side of Eq. \eqref{FinalRhos}.

In Ref. \cite{Me24NCevo}, we analyzed equation \eqref{FinalRhos} without the summation term on the left-hand side, which describes smooth evolution in phase space. There is only a time derivative on the left and a transition term on the right. From the transition term, we deduced the Boltzmann collision integral for two particular cases: electron-phonon interaction and three-phonon processes. While these two cases are probably the most common in solid-state theory, we cannot use them for this manuscript. The problem is that the whole treatment of smooth evolution in Sections 2 and 3 is given for massive particles that cannot be absorbed or emitted as a result of interaction. Also, we still cannot derive the collision integral in the general case. So here we want to give two more examples. These would be pair collisions of massive fermions and bosons, respectively.

The fermionic case is much simpler, so we start with it. We consider the mean number of particles in single-particle state 1, affected by a collision with a particle in single-particle state 2, which makes them go to single-particle states 3 and 4. We also consider the reverse process, so overall we have transitions between states $ \langle 1, 1, 0,0 \rangle  \leftrightarrow  \langle 0, 0, 1, 1\rangle$. Following the convention introduced in  \cite{Me24NCevo}, we denote non-coherent states with angular brackets on both sides. The relevant terms of the Hamiltonian are $V \hat{c}_1 \hat{c}_2 \hat{c}_3^\dagger \hat{c}_4^\dagger + V^* \hat{c}_1^\dagger \hat{c}_2^\dagger \hat{c}_3 \hat{c}_4$. We denote $q = |V|^2$. According to equation \eqref{Fermi}, the time derivatives of the probability of the system to be in each of the two considered states are given by
\begin{align}
\dot P_{\langle 1, 1, 0, 0 \rangle} = -q P_{\langle 1, 1, 0, 0 \rangle} + q P_{\langle 0, 0, 1, 1 \rangle} \nonumber \\
\dot P_{\langle 0, 0, 1, 1 \rangle} = -q P_{\langle 0, 0, 1, 1 \rangle} + q P_{\langle 1, 1, 0, 0 \rangle} .
\label{Dynamics}
\end{align}
To transition from probabilities of many particle state to probabilities of single-particle-states we use factorisation hypothesis, namely
\begin{align}
P_{\langle 1, 1, 0, 0 \rangle} =  P_{\langle n_1 = 1 \rangle} P_{\langle n_2 = 1 \rangle}  P_{\langle  n_3 = 0 \rangle}  P_{\langle n_4 = 0 \rangle} .
\label{Factorisation}
\end{align}
Becouse a single-particle state can not be occupied by more than one fermion, it is true that $P_{\langle  n = 0 \rangle} =  1 - P_{\langle  n = 1 \rangle}$, and the mean number of particles just coincide with probability $f_n = P_{\langle  n = 1 \rangle}$. We substitute these formulae and factorisation \eqref{Factorisation} into expression \eqref{Dynamics} we get 
\begin{align}
\dot f_1 = -q f_1 f_2 (1-f_3) (1-f_4) + q (1-f_1) (1 - f_2) f_3 f_4.
\label{Dynamics}
\end{align}
Summation over set of all permitted single-particle states $2, 3, 4$ yields the full collision integral.

Calculation for bosons is much more complicated since there can be arbitrary number of particles occupying a single-particle-state. For mean number of particles we have:
\begin{equation}
N_1 = \sum_{n_1, n_2, n_3, n_4} n_1 P_{\langle n_1, n_2, n_3, n_4\rangle}.
\label{Average}
\end{equation}

Transitions under consideration are
 $ \langle n_1-1, n_2-1, n_3+1, n_4+2 \rangle  \leftrightarrow  \langle n_1, n_2, n_3, n_4\rangle \leftrightarrow \langle n_1+1, n_2+1, n_3-1, n_4-1\rangle$. With formula \eqref{Fermi} we have:
\begin{align}
\dot P_{\langle n_1, n_2, n_3, n_4\rangle} = q n_1 n_2 (n_3+1)(n_4+1) P_{\langle n_1-1, n_2-1, n_3+1, n_4+1\rangle} + q (n_1+1) (n_2+1) n_3 n_4 P_{\langle n_1+1, n_2+1, n_3-1, n_4-1 \rangle} - \nonumber \\
 -q \left(  n_1 n_2 (n_3+1) (n_4+1) +  (n_1+1) (n_2+1) n_3 n_4 \right) P_{\langle n_1, n_2, n_3, n_4\rangle} .
\label{4BosonStart}
\end{align}
We take the time derivative of Eq. \eqref{Average} and substitute Eq. \eqref{4BosonStart} into it to get
\begin{align}
\dot N_1 = q \sum_{n_1, n_2, n_3, n_4} n_1 [  n_1 n_2 (n_3+1) (n_4+1) P_{\langle n_1-1, n_2-1, n_3+1, n_4+1\rangle} + (n_1+1) (n_2+1) n_3 n_4 P_{\langle n_1+1, n_2+1, n_3-1, n_4-1 \rangle} - \nonumber \\
 \left(  n_1 n_2 (n_3+1) (n_4+1) +  (n_1+1) (n_2+1) n_3 n_4 \right) P_{\langle n_1, n_2, n_3, n_4\rangle}  ].
\label{rate1}
\end{align}
We shift indices $n_2, n_3, n_4$ so that summation index would coincide with index in probability, and multiplier before probability gets shifted accordingly:
\begin{align}
\dot N_1 = q \sum_{n_1, n_2, n_3, n_4} n_1 [  n_1 (n_2+1) n_3 n_4 P_{\langle n_1-1, n_2, n_3, n_4\rangle} + (n_1+1) n_2 (n_3+1) (n_4+1) P_{\langle n_1+1, n_2, n_3, n_4 \rangle} - \dots ].
\label{rate2}
\end{align}

We make factorisation assumption and sum over $n_2, n_3, n_4$ using formula \eqref{Average} to get
\begin{align}
\dot N_1 = q \sum_{n_1} n_1 [  n_1 (N_2+1) N_3 N_4 P_{\langle n_1-1 \rangle} + (n_1+1) N_2 (N_3+1) (N_4+1) P_{\langle n_1+1 \rangle} - \nonumber \\
 \left(  n_1 N_2 (N_3+1) (N_4+1) +  (n_1+1) (N_2+1) N_3 N_4 \right) P_{\langle n_1 \rangle}  ].
\label{rate3}
\end{align}
We regroup summands, take $n_1$ into brackets, shift indices and obtain
\begin{align}
\dot N_1 = q \sum_{n_1} [ (N_2+1) N_3 N_4\left(  n_1^2 P_{\langle n_1-1 \rangle} -  n_1 (n_1+1)  P_{\langle n_1 \rangle}  \right)  +
 N_2 (N_3+1) (N_4+1) \left(  n_1 (n_1+1) P_{\langle n_1+1 \rangle} -  n_1^2  P_{\langle n_1 \rangle}  \right)  ].
\label{rate4}
\end{align}
Summation over $n_1$  yeilds
\begin{align}
\dot N_1 =  q (N_1+1)  (N_2+1) N_3 N_4 - q N_1 N_2 (N_3+1) (N_4+1)
\label{BosFin}
\end{align}
Summation over all possible combinations of all permitted single-particle states $2, 3, 4$ gives the collision integral.

Combining this result with the non-collisional part, and using $p/m = v$, $-\text{grad}_x U = F$, yields the full Boltzmann equation that contains diffusive, force, and collisional terms:
\begin{equation}
\dot f(\vec x, \vec v) + \vec v \cdot \text{grad}_x  f  + \vec F \cdot  \text{grad}_v  f  = \text{St} (f).
\label{FullBoltzmann}
\end{equation}
Thus, we have achieved the goal of the manuscript.

At the end, we want to note two things. First, we have used the factorization assumption without justification, just because it seems plausible and leads to the right result. However, in our derivation, already equation \eqref{FinalRhos} is irreversible, which has nothing to do with factorization. In some sense, this makes the presented investigation complementary to the research line that derives factorization \cite{Deng2024Hil6, Deng2025Hil6} (but, in our opinion, does not explain irreversibility).

Second, the form of the collision integral \eqref{BosFin} is usually justified just by saying "the probability of a boson leaving the initial state is proportional to the number of bosons, and because of stimulated emission, the probability of a boson coming to a state is proportional to the number of bosons already in this state plus one" \cite{LandauV10}. While this is true, $N$ represents the mean number of particles, and the contribution to the mean number from transitions to a state with $n$ particles is also proportional to $n$, see expression \eqref{Average}. Combining these two factors leads to terms proportional to $n^2$ in the intermediate expressions (\ref{rate3}, \ref{rate4}). These terms, however, cancel out before the final formula \eqref{BosFin}. It is interesting that it is not even true that the first term in \eqref{BosFin} corresponds to bosons transitioning to single-particle state 1 while the second term corresponds to bosons leaving single-particle state 1. In fact, the first term corresponds to transitions $ \langle n_1-1, n_2-1, n_3+1, n_4+2 \rangle  \rightarrow  \langle n_1, n_2, n_3, n_4\rangle \rightarrow  \langle n_1+1, n_2+1, n_3-1, n_4-1\rangle$, and the second term to transitions $ \langle n_1-1, n_2-1, n_3+1, n_4+2 \rangle \leftarrow \langle n_1, n_2, n_3, n_4\rangle \leftarrow \langle n_1+1, n_2+1, n_3-1, n_4-1\rangle$. Both terms contain both bosons leaving and coming to the single-particle state 1. These combinations lead to cancellation of terms proportional to $n^2$. Here, we have only shown this cancellation for the particular model; deriving this cancellation in general remains a demanding task.

\section{Summary and discussion}

It is typically thought that the Boltzmann equation is derived from classical mechanics, and it is assumed that classical particles are classical because they are considered “heavy”. Meanwhile, the Boltzmann equation is successfully used in the description of kinetic phenomena in solid states \cite{Ziman1960, LandauV10}, where the particles are quasiparticles, are described by quantum distribution functions (Bose–Einstein and Fermi–Dirac), and do not possess momentum, but rather quasimomentum. Concerning masses, electrons and holes typically have effective masses that are a fraction of the electron mass \cite{Anselm1981}, which is very “light”, and phonons are massless quasiparticles.

Yet thinking about them as point particles that are, at least partly, described by classical mechanics provides fruitful physical insights into transport phenomena in crystals. For example, the theoretical prediction of the properties of a field-effect transistor is based on the diffusion equation \cite{Anselm1981}, which more or less explicitly assumes the object of its description to behave like a classical gas. Such an application should have a proper justification.

The applicability of the mental model of a classical gas follows from the applicability of the Boltzmann equation. Its systematic derivation from quantum mechanics justifies the use of classical concepts for quasiparticles. A serious obstruction to this justification is the completely different mathematical methods that are used to derive the different parts of the Boltzmann equation.

The collision-integral part consists of instantaneous transitions of the system from one state to another, and the rate of such transitions, in the simplest case, is computed using the Fermi rule. The method employed by Dirac to derive the Fermi rule, which was used and extended in our previous work \cite{Me24NCevo}, was based on the smooth envelope ansatz \eqref{Psi}. We seek the wave function in the form $\Psi = \sum A \psi$, where $A$ is an amplitude slowly varying in time and $\psi$ is the wave function of an interactionless system.

The non-collision part of the Boltzmann equation is a consequence of the Liouville equation, which can be derived from classical mechanics. And we already know that classical mechanics can be derived from quantum mechanics. However, we see that the standard methods of derivation of classical mechanics (and, consequently, of the non-collisional part of the Boltzmann equation) are completely different from the method of derivation of the collision-integral part. Our goal was to derive the Boltzmann equation as a single piece from the Schrödinger equation, and to see where the line lies that separates the first part from the second part.

To accomplish this, we derive the Liouville equation here directly from Schrödinger’s equation, using the wavelet transform technique — the same one we previously used (without reference to wavelets) to describe quantum transitions. We see that fields that vary slowly, that is, with approximately the same rate as the amplitudes of the states, act on the system in such a way that it exhibits classical-like behavior, which is reversible and deterministic. By contrast, the interaction of the system with fast-oscillating fields results in behavior characterized by quantum transitions, which is probabilistic in nature and irreversible. We call the study of these two limits within one framework, and their distinction according to their qualitatively different behavior, the separation of scales.

The belief that the Boltzmann equation is derived through classical mechanics is known to lead to a very serious contradiction: classical mechanics is time-reversible, while the Boltzmann equation is irreversible. We carefully address the emergence of irreversibility in our theory \cite{Me24transmission, Me25Larotonda, Me24NCevo} and come to the conclusion that the source of irreversibility is non-coherence. When two waves overlap, non-coherence devours interference and produces irreversibility.

In the derivation presented in this manuscript, we have seen that the amplitudes satisfy the same equation \eqref{PreLiouville} as the probability densities, namely, the Liouville equation. This equation, by the famous Liouville theorem \cite{ArnoldMech}, conserves measure, which, when translated into simple language, means that different small parts of the envelope function, following their prescribed trajectories \eqref{Hamilton}, do not overlap. This guarantees that no interference arises, and that is why no irreversibility is present in the non-collisional part of the Boltzmann equation. This is in contrast with the collisional part, where the system can transition to one particular state from a handful of other states, and the amplitudes of these initial states interfere with each other.

This resolves the paradox of how the irreversible Boltzmann equation emerges from reversible classical mechanics. Our reasoning suggests that it does not. The non-collision part and the collision part are two different limits of quantum mechanics, which are the long-wavelength and short-wavelength limits, respectively.

Moreover, as we have noted in the introduction, the theory presented in this manuscript is an extension of the wave-packet theory. The wave packets, which represent the motion of a particle along a trajectory, are assumed to have a finite spectral width. In our theory of non-coherent evolution, which was presented earlier, we also assumed a finite spectral width of the states in order to explain why the reversible unitary evolution is replaced by irreversible stochastic evolution. We see that the two complementary parts of our theory rely on the same physical assumptions: the representability of the wave function as a composition of two parts (slowly oscillating and fast oscillating), and the finite spectral width of the states.

There is an alternative theory that explains the right-hand side of equation \eqref{FinalRhos}, which leads to the collision-integral part of the Boltzmann equation, known as quantum thermodynamics \cite{breuer2002openquant}. In this context, the right-hand side of equation \eqref{FinalRhos} is known as the Born–Markov approximation of the Lindblad master equation \cite{lindblad1976generators}. Quantum thermodynamics has successfully shown that a system described by the master equation tends to thermal equilibrium \cite{rigol2008thermalization, deutsch2010thermodynamic, dymarsky2018subsystem}. However, applying it in the context of physical kinetics seems problematic.

Quantum thermodynamics attributes the loss of coherence in a quantum system to its interaction with the environment. To obtain the correct form of the Lindblad equation, one must assume that the heat-bath relaxation time is small compared to the relaxation time of the system \cite{breuer2002openquant}. This essentially means that there is an additional interaction that is stronger than the interaction of interest. If such an interaction exists, it should result in a reduction of the particles' free path and thus contribute to the kinetic coefficients. However, in quantum thermodynamics this does not happen: the external interaction only affects the form of the evolution equation, making it irreversible, but does not affect its parameters. Our theory is free of such a contradiction.

In \cite{Me24NCevo}, we noted that our methods take into account only contributions from waves with wavelengths much smaller than $\Delta x$, the technical scale of integration (see Eq.~\eqref{smoothing} and the discussion there). Information about low-frequency components was thus lost, and we suggested that such low-frequency modes should be treated as the force term in the Boltzmann equation. Here, we have successfully provided such a description.

What is lacking in our derivation is a formal mathematical procedure for separating the potential into two parts, one of which would be “sharp” and contribute to the collision integral, while the other would be “smooth” and contribute to the force term. In Section~IV of the manuscript, we specifically required the interparticle interaction to be short-ranged, since our methods do not allow for a systematic treatment of long-range interaction potentials. Resolution of this problem would not only extend the results of this paper to a broader class of systems, including those with long-range interactions. A systematic exclusion of small-frequency contributions from integrals arising in perturbation theory is very important in cases where such integrals are divergent at zero frequency. This type of divergence is known as an infrared divergence. A formal separation of the potential into sharp and smooth parts, of which only the sharp part contributes to perturbation-theory integrals, would provide a physically clear and mathematically rigorous procedure for eliminating infrared divergences.

\acknowledgments{During the preparation of this work, the author used ChatGPT to improve the quality of English. It also helped with the literature search. After using this service, the author reviewed and edited the content as needed and takes full responsibility for the content of the publication.}

\bibliography{timerev, core}
\bibliographystyle{ieeetr}

\end{document}